\documentstyle[12pt,fleqn,psfig]{article} 
\newcommand{\ba}{\begin{array}}
\newcommand{\ea}{\end{array}}
\newcommand{\bd}{\begin{displaymath}}
\newcommand{\ed}{\end{displaymath}}
\newcommand{\be}{\begin{equation}}
\newcommand{\ee}{\end{equation}}
\newcommand{\bea}{\begin{eqnarray}}
\newcommand{\eea}{\end{eqnarray}}
\newcommand{\Rsl}{{\not \! \!{R}}}

\def\a{\alpha}
\def\b{\beta}

\def\l{\lambda}
\def\m{\mu}
\def\n{\nu}

\def\p{\pi}
\def\pr{\prime}

\begin{document}

\begin{flushright}
\tt{hep-ph/0105119} 
\end{flushright}
\vskip 5pt
\begin{center}
{\Large {\bf A new three flavor oscillation solution of the solar
neutrino deficit in $R$-parity violating supersymmetry}}
\vskip 20pt

\renewcommand{\thefootnote}{\fnsymbol{footnote}}

{\sf Rathin Adhikari $^{a,b,\!\!}$
\footnote{E-mail address: rathin@cubmb.ernet.in}},   
{\sf Arunansu Sil $^{a,\!\!}$
\footnote{E-mail address: arun@cubmb.ernet.in}}, 
and 
{\sf Amitava Raychaudhuri $^{a,\!\!}$
\footnote{E-mail address: amitava@cubmb.ernet.in}}  
\vskip 10pt  
$^a${\it Department of Physics, University of Calcutta, 92
Acharya Prafulla Chandra Road, \\ Kolkata 700009, India.}\\ 
$^b${\it Department of Physics, Jadavpur University,
Kolkata 700032, India.}\\
\vskip 15pt
{\bf Abstract}
\end{center}

{\small 
We present a solution of the solar neutrino deficit using three
flavors of neutrinos within  the R-parity non-conserving
supersymmetric model.  In vacuum, mass and mixing is restricted
to the $\nu_\mu$-$\nu_\tau$ sector only, which we choose in
consistency with the requirements of the atmospheric neutrino
anomaly. The $\nu_e$ is massless and unmixed. The flavor changing
and flavor diagonal neutral currents present in the model and an
energy-dependent resonance-induced $\nu_e$-$\nu_\mu$  mixing in
the sun  result in the new solution to the solar neutrino
problem.  The best fit to the solar neutrino rates and spectrum
(1258-day $SK$ data) requires a mass square difference $\sim
10^{-5}$ eV$^2$ in vacuum between the two lightest neutrinos.
This solution cannot accommodate a significant day-night effect
for solar neutrinos.}

\vskip 20pt

\begin{center}
PACS NO. 14.60.Pq, 26.65.+t, 12.60.Jv, 13.15.+g 
\end{center}


\vskip 30pt

\renewcommand{\thesection}{\Roman{section}}
\renewcommand{\thefootnote}{\arabic{footnote}}
\setcounter{footnote}{0}

Neutrino oscillation is  the most popular solution  of the solar
neutrino problem \cite{bahc,solar1} and the atmospheric
neutrino anomaly \cite{atm}.  Oscillation in vacuum or in matter,
through the MSW resonance mechanism, posits that neutrinos have
non-vanishing, non-degenerate masses  and that the basis defined
by these eigenstates does not coincide with the flavor basis.

Supersymmetry (SUSY) with $R$-parity non-conservation is an
extension of the Standard Model (SM) which is consistent with all
particle physics experiments and is phenomenologically rich
\cite{review}. It carries within it new interactions between
leptons and quarks which violate baryon ($B$) and lepton ($L$)
number. In this work we show that the flavor changing neutral
currents (FCNC) and flavor diagonal neutral currents (FDNC) due
to the $L$-violating interactions induce mixing amongst neutrinos
in matter, the key feature in this alternative solution of the
solar neutrino discrepancy, even though,  in vacuum, the $\nu_e$
state is massless and does not mix with the other neutrinos. We
also indicate how in this model the parameters can be chosen to
address the atmospheric neutrino anomaly, solving the solar
neutrino problem at the same time.

Origins of neutrino oscillation other than mass-mixing, notable
among them being non-standard interactions of neutrinos with
matter, like FCNC, were examined by Wolfenstein \cite{wolf}.  It
is noteworthy that FCNC and FDNC interactions can drive neutrino
oscillations even for massless neutrinos without any vacuum
mixing through an energy-independent resonance effect
\cite{masi}.  However, these solutions require large
$L$-violating couplings  near their present experimental upper
bounds \cite{guzzo,atf}.  This has been examined earlier in
connection with the solar \cite{masi,guzzo,bbk} and atmospheric
neutrino data \cite{atf,atf1} in the {\em two} flavor oscillation
framework.  The new explanation of the solar neutrino deficit
that we propose, in contrast, relies on an interplay between the
$\Rsl$-interactions for {\em three} flavors of neutrinos with
matter and their masses, keeping {\em vacuum mixing restricted to
the $\nu_\mu-\nu_\tau$ sector only}.

Imposing baryon number conservation, we focus on the 
following $L$-violating terms in the superpotential:
\be
W={1 \over 2} \l_{ijk} \; L_i \, L_j \, E_k^c + \l_{ijk}^{\pr} \;
L_i \, Q_j \, D_k^c,
\label{eq:spot}
\ee
assuming that bilinear terms have been rotated away with
appropriate redefinition of superfields. Here, $i$,$j$, and $k$
are generation indices, $L$ and $Q$ are chiral superfields
containing left-handed lepton and quark doublets and $E$ and $D$
are chiral superfields containing right-handed charged-lepton and
$d$-quark singlets.  There are nine $\l$ (antisymmetric in
$(ij)$) and twenty-seven $\l^{\pr}$ couplings, only a few of which
will be relevant for this analysis.

The interaction of neutrinos with the electrons and $d$-quarks in
matter induces transitions (i) $\n_i + e \rightarrow \n_j +e$,
and (ii) $\n_i + d \rightarrow \n_j + d$. (i) can proceed {\em via}
$W$ and $Z$ exchange for $i=j$, as well as {\em via} $\l$
couplings for all $i,j$, while process (ii) is possible through
$\l^{\pr}$ couplings and squark exchange. Here we concentrate
only on the $\lambda$-induced contributions. 

The time evolution of the neutrino flavor eigenstates ($\n_i, \;
i = e, \m , \tau $) is governed by
\bea
H&=&H^0 + h^{matter}  
\nonumber \\ 
&=&\left(
\begin{array}{ccc} E & 
0& 0 \\  0 & E+ S_+ - T_1 & T_2 \\ 0   &   T_2
& E + S_+ + T_1  \end{array}
\right) + \left(  
\begin{array}{ccc}
R_{11} +A_1 - A_2 & 0 &R_{13}\\  0 &  -A_2  & 0  \\
R_{13} &   0 &  R_{33} - A_2  
\end{array}
\right), 
\label{eq:H}
\eea
where $S_{\pm} = (m^2_3 \pm m^2_2)/4E$, $T_1 = S_- \cos
2\theta_{23v}$, $T_2 = S_- \sin 2\theta_{23v}$, $A_1 =
\sqrt{2}G_F n_e $, $A_2 = G_F n_N/\sqrt{2}$, and $R_{ij} =
\l_{ik1}\l_{jk1}n_e/4{\tilde m}^2$. $E$ is the
neutrino energy and  $\theta_{23v}$ the vacuum mixing angle in
the $\n_{\mu}-\n_{\tau}$ sector. $n_N$ and $n_e$ are the neutron
and electron number densities in matter and ${\tilde m}$ is the
slepton mass.  $A_1$ and $A_2$ in $h^{matter}$ arise from SM charged
and neutral current interactions, respectively.  In vacuum,
$h^{matter} = 0$ and $H$ contains mixing only in the $\nu_\mu -
\nu_\tau$ sector. In $h^{matter}$, we choose\footnote{In view of the
antisymmetry of $\l_{ijk}$ in ($i,j$), in order to generate the
mixing of the $\nu_e$ with the other neutrinos we have to choose
$k$ = 2 or 3.  For the latter choice, mixings due to $\Rsl$
interactions are very small; for example, $\l_{131}\l_{23 1}$ is
highly constrained from $\mu \rightarrow 3 e$ decay
\cite{review}.}  $k=2$ in the matter-induced contributions
$R_{ij}$. For anti-neutrinos, the time evolution is determined by
a similar total hamiltonian $\bar{H} = H^0 - h^{matter}$.

To obtain the mass eigenstates, first we rotate by $U^{\pr}  =
U_{23} U_{13}$ (where $U_{ij}$ is the standard rotation matrix)
and write the effective mass squared matrix, ${\widetilde{M}^2 \over
2 E } = H - E - A_1 - A_2$,  in the new basis as
\be
{{\widetilde {M}}^2 \over 2 E } \approx \left(
\begin{array}{ccc} 
R_{11} c_{13}^2- 2 R_{13} c_{23} s_{13} c_{13} + \Lambda_+ s_{13}^2 &
-R_{13} s_{23} c_{13} & 0  \nonumber \\ -R_{13} s_{23} c_{13} &
\Lambda_- & -R_{13} s_{23} s_{13} \nonumber \\ 0   & -R_{13} s_{23}
s_{13} & R_{11} s_{13}^2 + 2 R_{13} c_{23} s_{13} c_{13} +
\Lambda_+ c_{13}^2 
\end{array}
\right), 
\label{eq:mass1}
\ee
where
\be
\Lambda_\pm = \left[S_+ - A_1 + \frac{R_{33}}{2}\right] \pm  \left[S_-
\cos 2(\theta_{23\nu} - \theta_{23}) + \frac{R_{33}}{2}\cos
2\theta_{23}\right]
\ee
and $c_{ij} \equiv \cos \theta_{ij}$ and  $s_{ij} \equiv \sin 
\theta_{ij}$.
Furthermore, 
\bea
\tan 2 \theta_{23} =  2T_2/(2T_1 + R_{33});\;\;\; \tan 2
\theta_{13}= 2 R_{13} c_{23}/D_1;\;\;\; D_1 = \Lambda_+ - R_{11}.
\label{eq:mixang1}
\eea
Note that $\theta_{23} \approx \theta_{23v}$ while\footnote{This
follows as $\Lambda_+ -R_{11} \sim m_3^2/(2 E)$ is very large with
respect to $R_{13}$ in the sun.} $\theta_{13}\approx 0$ except near a
possible resonance, when $D_1 = 0$. We show below that this resonance
condition cannot be achieved in the sun. Consequently, to a good
approximation,  the third state in this basis decouples in eq.
(\ref{eq:mass1}). The upper left $2\times 2$ block is readily
diagonalised, resulting in three effective masses ${\tilde{m}}_i$ as:
\bea
{\tilde {m}}_1^2/(2 E)
=c_{12}^2 \left( R_{11} c_{13}^2  - R_{13} c_{23}  \sin 2\theta_{13} + \Lambda_+ s_{13}^2
 \right) + R_{13}
s_{23} c_{13} \sin 2 \theta_{12}  + \Lambda_- s_{12}^2 
\nonumber
\eea
\bea
{\tilde {m}}_2^2/(2 E)= s_{12}^2 \left(
R_{11} c_{13}^2  - R_{13} c_{23}  \sin 2\theta_{13} + \Lambda_+ s_{13}^2
  \right) - R_{13}
s_{23} c_{13} \sin 2 \theta_{12}  + \Lambda_- c_{12}^2 
\nonumber
\eea
\bea
{\tilde {m}}_3^2/(2 E)= 
R_{11} s_{13}^2 + R_{13} c_{23} \sin 2\theta_{13} + \Lambda_+ c_{13}^2,
\label{eq:eigen1}
\eea
where 
\be
\tan 2 \theta_{12} = {-2 R_{13} s_{23} c_{13} \over D_2};\;\; D_2 =
\Lambda_- - R_{11} c_{13}^2 + R_{13} c_{23} \sin 2 \theta_{13} -\Lambda_+
s_{13}^2.
\label{eq:mixang2}
\ee
A resonant enhancement of $\theta_{12}$ occurs when $D_2 = 0$. 

The  neutrino flavor eigenstates $\nu_{\alpha} = \nu_{e, \mu,
\tau}$ are related to the mass eigenstates $\nu_i = \nu_{1,2,3} $
by
\be
\nu_{\alpha} =    \sum_{i} U_{\a i} \; \n_i,
\label{eq:massflav}
\ee
\noindent
where $U_{\a i}$ are elements of the unitary mixing matrix
\bea
U&=&\left(
\begin{array}{ccc} c_{12} c_{13}&s_{12} c_{13} 
& s_{13} 
\\
-s_{12} c_{23}  -c_{12} s_{23} s_{13}
  & c_{12} c_{23}  - s_{12}
s_{23} s_{13}& s_{23} c_{13}
\\
s_{12} s_{23}  - c_{12} c_{23}    s_{13}& - c_{12} s_{23} -
s_{12} c_{23}       s_{13} &   c_{23} c_{13}
\end{array}
\right).
\label{eq:tmix}
\eea
We have chosen real $L$ violating couplings and as such there is
no $CP$ violating phase in the above mixing matrix.  Further, in
order to satisfy $0 \leq \theta_{12} \leq \pi/2$ in eq.
(\ref{eq:tmix}) for convenience, we take $\l_{121} \l_{321} <
0$. 

As noted above, level crossings and resonance behavior, which  are
energy dependent due to neutrino masses, can occur in two situations,
namely, (a) when $D_1 = 0$, and (b) when $D_2 = 0$.  Of these, only the
latter can be satisfied inside the sun, as we now discuss.  The sub-GeV
and multi-GeV zenith angle dependence of atmospheric neutrinos as well
as the energy dependence of the up-down asymmetry require $\Delta
m_{32} \approx m^2_3 \approx 10^{-3}$ eV$^2$ with maximal vacuum mixing
in the $\n_{\mu}-\n_{\tau}$ sector \cite{atm}.  The presence of $L$
violating interactions does not alter this significantly (see later).
On the other hand, $n_e$ at the core of the sun is about $1.13 \times
10^{12}$ eV$^3$. Thus even for $E$ as high as 20 MeV, it is not
possible to satisfy the (a) resonance condition and hence we consider
only the (b) resonance in the subsequent discussion of the solar
neutrino data. At resonance, $\theta_{12} = \pi/4$, while the
other mixing angles are $\theta_{13} \sim 0$ and $\theta_{23} =
\theta_{23v}$.  Recall that away from resonance,  $\theta_{12}
\sim 0$ and for vacuum propagation only $\theta_{23} =
\theta_{23v}$ is non-zero in eq. (\ref{eq:tmix}). At first
glance, one might think that if $U_{13}$ in vacuum is very small
then solar neutrinos will be almost unaffected by the mass of
$\n_{\tau}$ and analysis with three neutrino flavors may not be
essential.  However, unlike in the SM, where only $\n_e$
interactions with matter are relevant for neutrino oscillation,
in the $\Rsl$ supersymmetric Model, FCNC and FDNC interactions of
all three flavors of neutrinos turn out to be important. In fact,
one can see from eqs.  (\ref{eq:H}) and (\ref{eq:mixang2}) that
$R_{13}$, arising from FCNC interactions, appears in $\tan 2
\theta_{12}$ and plays a pivotal role.

We now turn to the oscillation of solar neutrinos due to their
interaction with matter inside the sun.  As already discussed,
$\n_e$ in the sun can experience only one of the two resonances.
$s_{13}$ in eq. (\ref{eq:tmix}) is very small as noted
earlier and we  use the survival probability of $\n_e$ valid
for a two flavor analysis:
\bea
P_{{\n}_e \rightarrow {\n}_e}= {1 \over 2}+ \left( {1 \over 2} -
P_{jump} \right) \cos 2 \theta_{12} (x_1) \cos 2 \theta_{12}
(x_0),
\label{eq:prob}
\eea
where $x_0$ is the production point inside the sun and $x_1$ the
detection point at earth\footnote{Notice that $\cos 2\theta_{12}(x_1) =
1$, corresponding to $\theta_{12} = 0$ in vacuum.}.  The jump
probability is $P_{jump} \approx \exp[- \pi \gamma_{res} F /2 ]$,
$\gamma_{res}$ being the adiabaticity parameter.  $F =1$ for the
exponential density profile since the vacuum mixing angle  is zero and
\bea
\gamma_{res} =  {{\tilde {m}}_2^2 -   {\tilde {m}}_1^2 \over 4 E
\dot{\theta}_{12} } \simeq \frac{m_2^2}{E} \left(\frac{p}{\kappa}\right)^2
\left( {n_e \over \dot{n}_e} \right)_{res},
\label{eq:adia}
\eea
where $\kappa = \left(2 \l_{121}^2 - \l_{321}^2 \right) /
(8 {\tilde{m}_{\mu}}^2) +  \sqrt{2} G_F   \sim \sqrt{2} G_F$
\footnote{$\l_{121}$ and $\l_{321}$ are tightly constrained \cite{review}.
Besides significant cancellation between these terms is possible if
they are of same order.} and $p = |{\l_{121} \l_{321}\over 4 m_{\tilde
\mu}^2}|$.

In order to obtain the best-fit values of $\Delta m_{12} \approx
m_2^2$ and $p$, we have performed a $\chi^2$ analysis using the
Standard Solar Model (SSM) \cite{bah} and the solar neutrino
rates from the Homestake ($Cl$), Gallex, Sage, and Kamiokande
($K$) experiments \cite{solar1}.  We have also used the latest
$SK$ rates and spectrum data for 1258 days \cite{SK1258}. Taking
into account the production point distributions of neutrinos from
the different reactions ({\em e.g.}, $pp$, $pep$, $^7Be$, $^8B$
{\em etc.})\footnote{We have dropped a small contribution 
from the $hep$ process.}, we have calculated the averaged
survival probabilities using eq. (\ref{eq:prob}). Here, we
include a parameter $X_B$ to take into account a possible
deviation of the overall normalization of the $^8B$ flux from its
SSM value. We set $\theta_{23v} = \pi /4$. The best-fit values of
the parameters are presented in Table 1 along with
$\chi^2_{min}$, the goodness of fit ($gof$), and the calculated
rates using these values of the parameters. Case (1) is a fit to
the total rates.  Note that the best-fit parameters result in an
unusually good fit to the $Cl$ rate and the $Ga$ prediction is
right near the average of the Sage and Gallex data.  We have
found that the fit improves even more if the $K$-rate is
excluded. In case (2) we have fitted the $SK$ spectrum while (3)
is a fit to the total rates and the $SK$ spectrum\footnote{We
have checked that the fit (3) is essentially unchanged if the
$SK$ rate is excluded from the fit.}. In Fig.  1 is shown the
calculated spectrum for $SK$ for the best-fit parameters along
with the experimental data.  Also shown is the charged current
spectrum expected at SNO for one sample case, the best-fit values
in case (3)\footnote{For these best-fit values, the prediction
for the neutral and charged current rates at SNO, normalized to
the SSM, are 0.56 and 0.44, respectively.}.

\begin{center}
\begin{tabular}{|c|c|c|c|c|c|c|c|c|c|c|}
\hline
&\multicolumn{5}{|c|}{Best-fit
Values}&\multicolumn{5}{|c|}{Corresponding Rates}\\ \cline{2-11}
&$p$&$ m_2^2$&$$ & & $\chi^2$ & $Cl$
& $Gallex$ & $Sage$ & $K$ & $SK$ \\ 
Case&$(10^{-24}$&$(10^{-5}$&$X_B$&$dof$&$(gof)$ & ($0.33 \pm$
&($0.52 \pm$ &($0.60 \pm$ &($0.54 \pm$ &($0.451 \pm$ \\
&eV$^{-2}$)&eV$^2$)&& &&$0.029$)&$0.06$)&$0.06$)&$
0.07$)&$ 0.016$) \\
\hline
1&$0.595$&$1.063 $&0.845 & 2 & 1.71 & 0.326
& 0.561 & 0.561 & 0.478 & 0.455 \\
& & & & & (42.5) & & & & & \\
\hline
2&0.009&0.01&0.446&16 &18.76&0.582&0.947&0.947&0.446&0.446\\
& & & & & (28.1) & & & & & \\
\hline
3&0.360&0.980&0.560 & 21 & 25.53 & 0.364 &
0.558 & 0.558 & 0.456 & 0.447 \\
& & & & & (22.5) & & & & & \\
\hline
\end{tabular}
\end{center}
\begin{description}
\item{\small \sf Table 1:} {\small \sf  The best-fit values of
the parameters, $p= |{\l_{121} \l_{321} \over 4 m_{\tilde
\mu}^2}|$, $m_2^2$, and $ X_{B}$ from fits to (1) all
rates, (2) the $SK$ spectrum, and (3) rates and $SK$ spectrum.
The rates for the different experiments obtained using these
best-fit parameters are also shown.}
\end{description}

The best-fit values of $\Rsl$ couplings in Table 1 are
consistent with the existing constraints. For example, in case
(3), choosing $m_{\tilde\mu} \sim 100$ GeV, we get $\l_{121}
\l_{321} \approx 0.0144$.  $\l_{121}$ is constrained from $\mu
\rightarrow e \bar{\n_e} \n_{\mu}$ decay (with selectron exchange tree
level diagram apart from the SM $W$ exchange diagram).  The
bound on $\l_{321}$ is from $R \equiv \Gamma(\tau
\rightarrow e \n \bar{\n}) /\Gamma(\tau \rightarrow \mu \n
\bar{\n})$ which gets a contribution from a selectron exchange
diagram. For $m_{\tilde e} \sim 200$ GeV or more, the
requirements are easily satisfied \cite{review}. 

Turning now to atmospheric neutrinos \cite{atm}, for a
simple-minded analysis we can consider the earth to be a slab of
a single density. $n_e$ in earth lies in the range $(3-6) N_A$
cm$^{-3}$. So the resonance condition, $D_2 = 0$, cannot be met
for atmospheric neutrinos having energy near the GeV range. In
order to explain the observed zenith angle dependence, we must
choose $\Delta m_{32} \sim 10^{-3}$ eV$^2$.  This precludes the
occurrence of the other resonance, $D_1 = 0$. Since neither
resonance condition can be satisfied, there will be almost no
effect on atmospheric neutrino oscillation due to the $L$
violating interactions as the associated couplings are very
small. So one can consider the mixing matrix in eq.
(\ref{eq:tmix}) valid for vacuum for which only $\theta_{23v}$ is
non-zero. Thus the solution to the atmospheric neutrino anomaly
is just the standard two neutrino mass-mixing one.

The neutrino masses and mixing pattern in vacuum required in this
solution can naturally arise in many models. For example, the
trilinear couplings in eq. (\ref{eq:spot}) contribute to the
neutrino mass matrix at the one-loop level through slepton or
squark  exchange diagrams \cite{hall}.  In particular, from the
$\l^{\pr}$ couplings one obtains:
\be
m_{ij}^{loop}= {3 \, m_b^2 \; ( A_b + \m \tan \b \, ) \over 8 \p^2 \;  
{\tilde
m_b}^2} \; \l^{\pr}_{i33} \, \l^{\pr}_{j33}.
\label{eq:mrpar}
\ee
where $A_b$ and $\m$ are soft SUSY-breaking parameters, ${\tilde
m_b}$ is the $b$-squark mass and $\tan\b$ is the ratio of two
Higgs vacuum expectation values. The last two generation indices
in $\l^{\pr}$ have been chosen as 3 for which the loop
contributions are enhanced {\em via} the $b-$quark mass.  We
remark that $m_{ij}$ is very small when  $i =  1$ and/or $j = 1$
because of the more stringent constraint \cite{review} on
$\l^{\pr}_{133}$. Notice that this mass matrix can correspond to
almost maximal mixing for $\n_{\m}$ and $\n_{\tau}$ if
$\l^{\pr}_{233} \approx \l^{\pr}_{333}$, with two neutrino masses
very small and one neutrino having significantly higher mass
$m_3 \approx 2 \; m_{33}^{loop}$, which can be suitably chosen by
taking appropriate values of the different parameters in
(\ref{eq:mrpar}).  It should be borne in mind that $m_2$ depends
on the difference of $\l^{\pr}_{233}$ and $\l^{\pr}_{333}$ and
can be several orders less than the mass of the heavier neutrino
while there will be almost maximal mixing.  The remaining
neutrino mass is $m_1 \approx 0$. Thus masses and vacuum mixings
can be as required in the model under consideration.

This neutrino mixing pattern also satisfies the bound ${U_{13}}^2
\leq 0.04$ in vacuum from the CHOOZ reactor experiment
\cite{chooz}. In fact, in vacuum ${U_{13}}^2 = 0$.

A comment about the earth regeneration effect for solar neutrinos
is pertinent. The $\nu_e$ is unmixed with the other neutrinos in
vacuum. As $n_e$ in earth is about two orders less than that near
the core of the sun,  no resonance condition will be
satisfied\footnote{Except for neutrinos with $E > 10$ MeV
passing very near the centre of the earth.}. Hence,  there will
not be an earth effect for solar neutrinos. In comparison with
the small angle MSW fits \cite{barg1}, the somewhat larger
best-fit $\Delta m_{12}$ and the zero value of $\theta_{12}$ in
vacuum here, result in a smaller day-night effect.

Though our discussion has been within the framework of $R$-parity
violating SUSY, there are other models \cite{otmod} where FCNC
and FDNC interactions are present. Our results can be adapted to
these scenarios in a straight-forward manner.

\vskip 10pt

\section*{Acknowledgements}
RA is supported by D.S.T., India, A.S. enjoys a fellowship from
U.G.C., India while AR has been supported in part by C.S.I.R.
and D.S.T., India.

\vskip 30pt
\begin{figure}[hb]
\vskip -1.00in
\psfig{figure=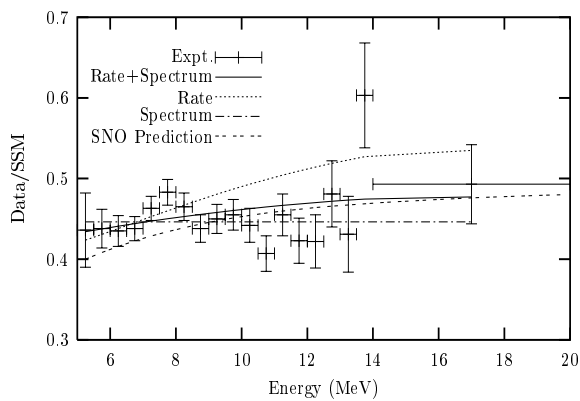,width=14.0cm,height=22.0cm} 
\vskip -5.10in
\caption{\sf \small   
The calculated SK solar neutrino spectrum for the best-fit
parameters $\Delta m_{12}$, $p$, and $X_{B}$ from (1) rates, (2)
SK spectrum, and  (3) rates and spectrum. The SK 1258-day data
\cite{SK1258} and the predicted SNO charged current
spectrum for the best-fit (3) are also shown.  } 
\end{figure}
\end{document}